# Global organization of the lexicon

M. Sigman [1, 2] and G.A. Cecchi [1, 3]

[1] *Laboratory of Mathematical Physics, Center for Studies in Physics and Biology and*

[2] *Laboratory of Neurobiology, The Rockefeller University, 1230 York Avenue, 10021 New York, NY.*

[3] *Functional Neuroimaging Laboratory, Department of Psychiatry, Weill Medical College of Cornell University, 1300 York Avenue, 10021 New York NY, USA*

**The lexicon consists of a set of word meanings and their semantic relationships. A systematic representation of the English lexicon based in psycholinguistic considerations has been put together in the database Wordnet in a long-term collaborative effort[1]. We present here a quantitative study of the graph structure of Wordnet in order to understand the global organization of the lexicon. We find that semantic links follow power-law, scale-invariant behaviors typical of self-organizing networks. Polysemy, the ambiguity of an individual word, can act as a link in the semantic network, relating the different meanings of a common word. Inclusion of polysemous links has a profound impact in the organization of the semantic graph, converting it into a *small world*, with clusters of high traffic (hubs) representing abstract concepts. Our results show that polysemy organizes the semantic graph in a**

**compact and categorical representation, and thus may explain the ubiquity of polysemy across languages.**

A pressing issue in linguistics, philosophy and brain science is the formal characterization of meaning, i.e. the formalization of our intuition of conceptual content. Language is a privileged window into the mind in its proposed double role of mediator and shaper of concepts[2] and lexical semantics, the mapping between word form and word meanings, is a testing ground for the problem of characterization of meaning within the domain of linguistics. Against the classical empiricist and reductionist interpretations of meaning as the character of the link of individual concepts and the external world, the holistic view proposes that mental concepts arise as an emergent property of their inter-relationships rather than as a property of their individual experiential correspondence [3,4]. The different relationships between meanings (antonomy, hypernymy, etc.) provide a first insight into the holistic nature of semantics. However, the binary nature of these relationships fails to describe longer-range semantic connections between word-concepts. As an example of how meanings can be related through long chains of semantic relationships, when the words "stripes" and "lion" are presented, one thinks of the word "tiger", establishing the trajectory lion-feline-tiger-stripes. Dictionaries also make evident the intrinsic holistic nature of languages, as all individual entries must be bootstrapped from other entries in a self-referential fashion. If meaning not only results from a correspondence with external objects, but also depends on the inter-relationships with other meanings, an understanding of the lexicon as a collective process implies a characterization of the structure of the graph, i.e. the global organization of the lexicon.

A word form is a label that identifies a meaning. However, meanings are not mapped to word-forms in a one to one fashion. Two words forms corresponding to the same meaning are said to be synonyms; a word form that corresponds to more than one meaning is said to be polysemous. All known languages are polysemous, but it is not yet clear whether the existence of polysemy is an historical accident, which an "ideal" language should avoid, or whether it may be important for abstract and metaphoric thought and imagery, and in general, for the organization of meanings[5].

We will investigate the statistics and organization of the following semantic relationships: antonymy, hypernymy (hyponymy), and meronymy (holonomy) (parentheses refer to inverse relationships). While antonymy is well known and needs no explanation, the other two relationships are less familiar. A hyponym is a meaning that acquires all the features of its hypernym, which is a more generic concept; for example, tree is hypernym of oak. Dictionaries make implicit use of the hypernym relationship by defining a word as its hypernym (which resumes all the basic features) and its specific attributes[1]. Meronymy is the relation of being part of; for example, branch is a meronym of tree. Meanings can also be related through common word forms (polysemy); for example, *a body of persons officially constituted for the transaction or superintendence of some particular business* and *a flat slab of wood* are two meanings that are related through the word form *board*. This relationship may seem arbitrary, accidentally linking unrelated meanings. In many cases, however, a chain linking the two different meanings may be established. For instance, the Oxford English Dictionary makes the relationship between the two meanings of the word *board* explicit: "A table at which a council is held; hence, a meeting of such a council round the table."

The lexicon then defines a graph, where the points are the different meanings and semantic relationships are the links. The vertices are nouns, adverbs, verbs or adjectives; in what follows we will present results based on an analysis of the set of nouns. Graph theory provides a number of indicators or measurements that characterize the structure of a graph: the statistical distribution of links - which gives an idea of the homogeneity and scaling properties of the graph -, the mean shortest distance between any two points of the graph - which gives an idea of its size, or diameter -, the clustering index - which provides a measure of the independency of neighboring links -, and the traffic - which measures the number of trajectories passing through each vertex, and so identifying the most active hubs. We will show several global properties of the set of nouns: a) all semantic relationships are scale invariant, typical of self-organizing graphs; while the semantic network is dominated by the hypernymy tree, which works as the skeleton of the set of nouns, the inclusion of polysemy produces a drastic global reorganization of the graph, namely b) it is converted into a *small world*[6,7], where all meanings are closer to each other, c) simplices (subgroups of fully connected meanings) become the regions of more traffic, and d) distances across the network are not in correspondence with their deepness in the hypernymy tree.

According to the latest version of Wordnet, the number of noun meanings is 66025 although this number is, of course, arbitrary and variable. In this work we will consider 4 types of relationships, hypernymy (I) antonymy (II) meronomy (III) and polysemy (IV). Antonomy and polysemy relationships are symmetric, while hypernymy and meronymy have hyponomy and holonomy as inverses. We will consider the set of hypernyms-holonomys as type I relationships, antonomy as type II relationships, meronomy-holonomy as type III relationships and polysemy as type IV relationships.

To address scaling in the network of meanings we computed the distribution of links (first-neighbour connections). Antonomy was excluded, because each meaning has 0 or 1, or rarely 2 antonyms. Remarkably, all relationships showed power-law behaviors reflecting scale invariance in this graph (Figure 1a). The relationship of hypernymy is the one with more links, relating at most 400 meanings. The word-form relationship relates up to 32 meanings through the most polysemous word *head*. The statistics of links provides a local measure to characterize a graph, of how uniform it is in the number of neighbors. Crystalline graphs, for example, have a fixed number of links. The power-law distribution of polysemy states that there are very few words that connect a high number of meanings, rendering them as particularly important ones. It has been shown that graphs with power-law distribution of links are very sensitive to the removal of these particular nodes [8]. Moreover, power-law distributions are a hallmark of self-organizing systems [9], reflecting the high quality of Wordnet despite the necessary arbitrariness with which it was built.

The statistics of links, however, does not fully characterize a graph; as seen in the toy model of Figure 1b, two graphs with equal distribution of links may be radically different. We therefore proceeded to calculate more global and characteristic measures of the graph: the characteristic length and the clustering. The characteristic length is the median of the minimal distance between pairs of vertices, and it essentially gives an idea of the diameter of the graph, measuring how far apart form each other two meanings typically are. Clustering is a measure of local structure, as it results from averaging the probability of two meanings being connected to each other given that they are both connected to a third common meaning. It is well known that social networks, for instance, are highly clustered: if A is friend of C, and B is friend of C, then A and B are also likely to be friends. Clusters define islands within the graphs, regions of very high internal connectivity.

We calculated the clustering and characteristic length for the 8 graphs that result from adding the different semantic relationships to the hypernymy tree (graph {I}). The graph {I,II} has all links of hypernymy and antonymy, the graph {I,IV} hypernymy and word-forms and son on. The graph {I,II,III,IV} contains all the possible links. All the graphs have the same number (66025) of vertices. Overall, as links are being added to a graph of fixed number of vertices, the characteristic length decreases and the clustering increases; however, the particular distribution of links may have very different effects on the length as well as the clustering of the graph[6]. To provide a measure of normalization we generated semi-random graphs which were constructed adding random links to the hypernymy graph. For each semantic graph we generated a corresponding semi-random graph with the same number of links.

Figure 2 summarizes the results for the 8 graphs. Overall, the impact of adding antonyms to the graph is very low. While this essentially results from the fact that there are very few (1849) antonym relationships, a small number of links may have a profound impact on the global organization of the graph[6]. The most important effect on the global structure of the hypernymy tree results from adding the polysemy relationship (Figure 2a) The characteristic length is reduced from 11.9 to 7.4 and the clustering increases from 0.0002 to 0.06. Graphs with random links are typically compact (i.e. have minimal characteristic length) and display very low clustering. As can be observed in the comparison with the semi-random graphs, the addition of the polysemy relationship compacts the graph as much as adding the same number of random links. At the same time the clustering is significantly increased, as demonstrated by the existence of vertices (green dots) with a high number of connected neighbors. In a way, it creates what is known as a small world, a clustered short-range graph. In comparison, adding the meronomy relationship has less of an effect in reducing the characteristic length, as indicated by the fact that adding the same number of random relationships produces a significantly smaller graph.

Given that polysemy has a profound impact on the clustering and characteristic length, we investigated the effect of polysemy in the relationship between mean distance and deepness in the hypernymy tree (minimal distance to the root) In the hypernymy graph, the way to go from one meaning to another is climbing up and down the tree. Thus, the mean distance from a node to the rest of the vertices, progresses as one goes deeper in the tree (Figure 3, blue). When polysemy is added, this correlation, while still present, becomes very weak (Figure 3, red).

A complementary measure of global organization in a graph is the *traffic* through any given vertex, counting the number of trajectories passing through it[10] (see methods). When polysemy is not present, different taxonomic groups are the meanings with more traffic. This appears to be an undesired artifact of the large number of related genera. In a tree structure the root is the center of traffic only if the branches are homogeneous; otherwise, highly connected child-vertices take over. The most active vertices in the graph I-II-II correspond to meanings with high number of hyponyms (Table I, Figure 4b blue dots), or which belong to heavy branches of the hypernymy tree (Figure 4c). When polysemy is added the distribution of vertices with high traffic forms clusters (Figure 4a, red dots) corresponding to central abstract meanings like *head* (red arrow) *line* (pink arrow) or *point* (orange arrow). Moreover, traffic becomes essentially independent of connectivity (Figure 4b, red dots). Interestingly, these three clusters correspond to the groups of meanings of the most polysemous words, respectively, *head* (p=30), *line* (p=29) and *point* (p=24). The vertices with highest traffic in the graphs with and without polysemy are non-overlapping (Figure 4c), showing a relation of orthogonality between polysemy and the other semantic relationships.

Different lines of research have provided evidence of mental navigation in the semantic network, implying that our observations of the global organization (and reorganization after the inclusion of polysemy) of the semantic network might be important for the understanding of neuronal representation of meanings[11-15]. It has been shown that the time to respond whether a statement linking two meanings is true or not depends on how far the two related meanings are in the hierarchy of hypernyms[16]. For example, it takes more time to respond true or false to the statement ``A canary has skin'' than to the statement ``A canary can fly'' which in its turn requires more time than the statement ``A canary sings''. Categorical effects in the retrieval of lexical knowledge have been also documented in lesion, as well as in imaging and electrophysiological approaches[17-21]. Word priming studies also provide evidence of semantic navigation; when subjects are asked to judge whether a string of letters is a word, judgements occur more rapidly following a semantically related item than an unrelated item[22]. Subliminal presentation of a word can also improve reaction times on a task related to a subsequently semantically related presented word[23,24]. An ambiguous (polysemous) word primes the different meanings associated with this word form[25]. Consequently, our study of the semantic network implies that global as well as local graph properties like connectivity, clustering, minimal distance and traffic should have psycholinguistic and neurobiological correlates. Specifically, we predict that the hubs of network traffic will have a statistical bias for priming, given the evidence for parallel activation of associated meanings.

Polysemy has remained a central challenge to artificial language[26]. With a one to one mapping of meanings to words, decoding meanings from word forms would result, of course, a trivial problem. Why are then languages polysemous? Our investigation of the semantic graphs including different number of relationships shows that addition of polysemy has well desired features in the global organization of the graph: it is converted into a small world, where meanings are typically closer to each other and central concepts (which are the most polysemous, and also the most familiar[27]) become the center of more trajectories, the hubs of the lexicon. This formalizes the concept that polysemy may be crucial for metaphoric thinking, imagery and generalization[5] and is in agreement with anatomical, lesion and functional studies that show that the right hemisphere is involved in linking ambiguous word meanings, metaphoric thinking, abstraction and in understanding ironic content[28-31].

Graph theory has been used to understand social and biochemical networks and the organization and dynamics of the World Wide Web [7,32-34]. Our results provide a quantitative framework to relate mental associations to the semantic structure of language, show that the semantic graph is scale invariant and suggest that the reorganization of the network after the inclusion of polysemy might be crucial for psycholinguistic considerations.

**Methods**

Grouping all semantic relationships in different classes is ultimately an arbitrary decision, although this grouping was based on a long history of research in psycholinguistics [1]. In WordNet, the relation of meronomy is actually a class of relations including meronomy of component (i.e. branch is component-meronomy of tree), member (tree/forest) and stuff (glass/bottle). Our study did not discriminate between different types of meronomy; another semantic relationship between nouns that we have not explored in this work is the relation of attribute. The minimal distance between vertices was computed adapting a publicly available version of Dijkstra's algorithm [10]. For a vertex $i$, the average minimal length was calculated by averaging the minimal distance from $i$ to all the other vertices in the graph. $\langle d \rangle_i = \frac{1}{66025} \sum_{j \in \{Menaings\}} dist_{min}(i,j)$ The characteristic length was computed as the median of the distribution of average minimal lengths across all vertices in the graph. Clustering was computed as the ratio of connected neighbours (*CN*) to the maximal possible number of connected neighbours, given by the formula $Clust = \frac{2 \cdot CN}{N*(N+1)}$ where *N* is the number of neighbours. All semantic graphs were obtained from the WordNet database. Semi-random graphs were generated as controls, by adding random links to the original hypernymy graph. A different number of random links were added to generate graphs of equivalent number of links, to control for the different graphs I-III I-IV and I-III-IV. Figure 2a shows an example of each of this graphs, to calculate the characteristic length and clustering of the semi-random graphs, we generated 7 different graphs for each condition and the averages are shown. Standard deviations were not shown because in all cases they were below 1%. The traffic was computed as the limit of the exponentiation of the graph; for an integer *N* and a graph *g*, $g^N_{ij}$ computes the number of trajectories of length *N* connecting vertices *i* and *j*. For $N \to \infty$, the exponentiation can be approximated by the largest eigenvalue $\lambda_1$ and its corresponding eigenvector. Because the composition of a trajectory with a loop gives a trajectory with the same path, but longer N, this limit is guaranteed to converge to the maximum number of paths and therefore $\lambda_1$ measures the limiting behavior of trajectories throughout the graph, i.e. the 'traffic' [10].

**Acknowledgements:**

We would like to thank Marcelo Magnasco for his guidance and support, Alex Backer and Sidarta Ribeiro for useful comments on the manuscript, and Michael Posner for helping us to relate our work to the psycholinguistics and cognitive literature.

Figures

1.- **Scale-invariant distribution of all semantic links.** A) Histogram showing the distribution of meanings as a function of number of links for the hypernymy-hyponymy (blue), polysemy (green) and meronymy-holonomy relationships. They all follow a power law (scale invariant) behavior, as seen by the linear dependence in the log-log plot. B) Cartoon of two graphs with identical distribution of links, but clearly distinct structure, showing that statistics of links is far from being a complete description of the shape of the graph.

2.- **Including polysemy results in a small world organization of the semantic graph**.
A) The histogram of average minimal length (see methods) and the plot of connected neighbors (*CN*) as a function of neighbors (*N*) is plotted for different graphs. The maximal number of *CN* is given by the formula $CN_{max} = \frac{N \cdot (N+1)}{2}$ which is plotted in the blue line. Points close to this line are highly clustered while the ones close to the horizontal axis are not. Since each graph has different number of links, for each one (except the hypernymy tree) we generated a control graph consisting of hypernymy + random links, to match the number of the original graphs. Data from the original plots are displayed in the left most columns, and data from the semi-random graphs in the 2 right most columns. The length and clustering are shown for the hypernymy tree (first row), hypernymy + meronomy (2nd row) hypernymy + polysemy (3rd row), hypernymy + meronomy + polysemy (4th row). As can be seen, adding word forms reduces the characteristic length as much as adding random relationships would do, but with a considerable increase of clustering (even for highly connected nodes) In addition, including polysemy adds vertices (green dots) of high connectivity and high clustering. Those correspond to cluster of meanings that are connected through a common word. b) Characteristic length for all possible graphs combining the hypernymy tree and the different semantic relationships (see text). As can be seen in the column corresponding to {I-IV}, adding polysemy, reduces the characteristic length as much as random relations do. c) Addition of semantic relationship on top of hypernymy increases the clustering of the graph. Thus polysemy results in a clustered a compact graph, a small world.

3 - **Correlation between mean distance and deepness in the hypernymy tree**. A scatter of average minimal distance versus depth in the hypernymy tree with and without the inclusion of polysemy. In the hypernymy tree (blue), distance scales with E with a tight correlation (r=0.902). After adding polysemy (red) the correlation is very weak (r=0.54) showing the hierarchy of hypernymy has low impact in distance between meanings in the full semantic graph.

4 –**Polysemy converts meanings related through the most polysemous words in the hubs of the graph**. A) The value of the first eigenvalue (a measure of traffic, see text and methods) for the graphs with all relations (red) and without polysemy (blue) as a function of distance (a) connectivity (b) and meaning index (c). a) The word forms produce clusters (marked with the arrows) which correspond to the most polisemous words: head (green arrow), point (pink arrow) and line (orange arrow). B) Traffic is independent of connectivity after the inclusion of polysemy (red dots), in the hypernymy tree, however, the points with mot hyponyms and thus the most connected ones become hubs. C) The value of the eigenvalue as a function of the meaning number according to the Wordnet (hypernymy based) order. As can be observed, the projections are almost orthogonal and the cluster (indicated with the red arrow) is spread in the order of hypernymy. The centers of high traffic occupy discrete positions corresponding to the heavy branches (the one with more meanings) in the hypernymy tree.

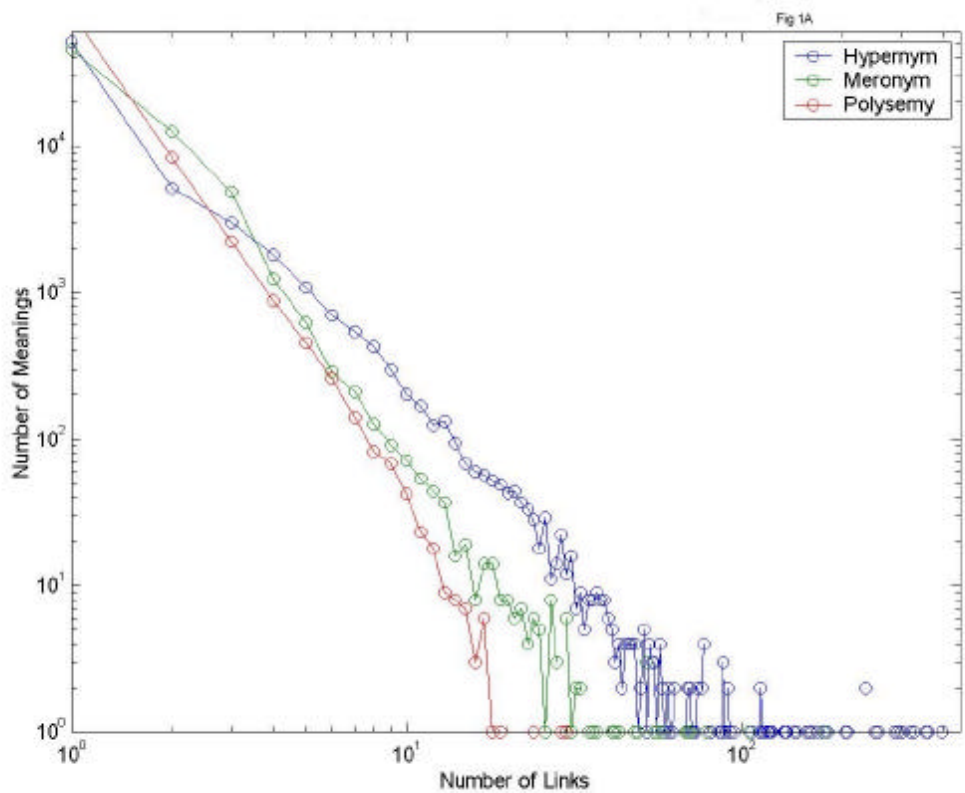

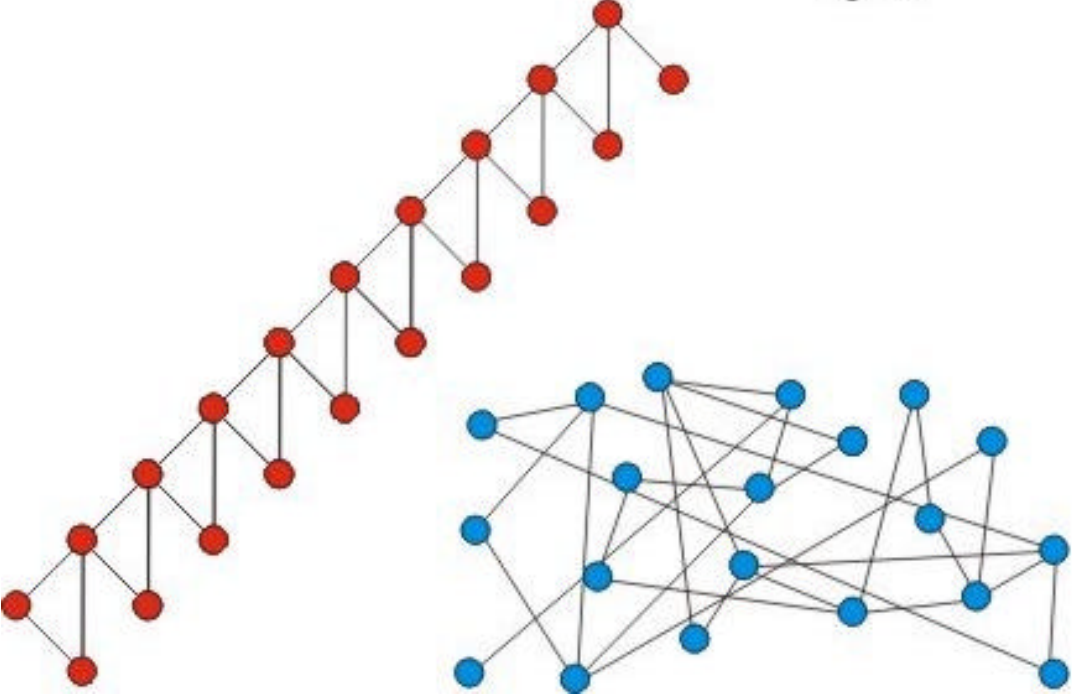

Fig 1B

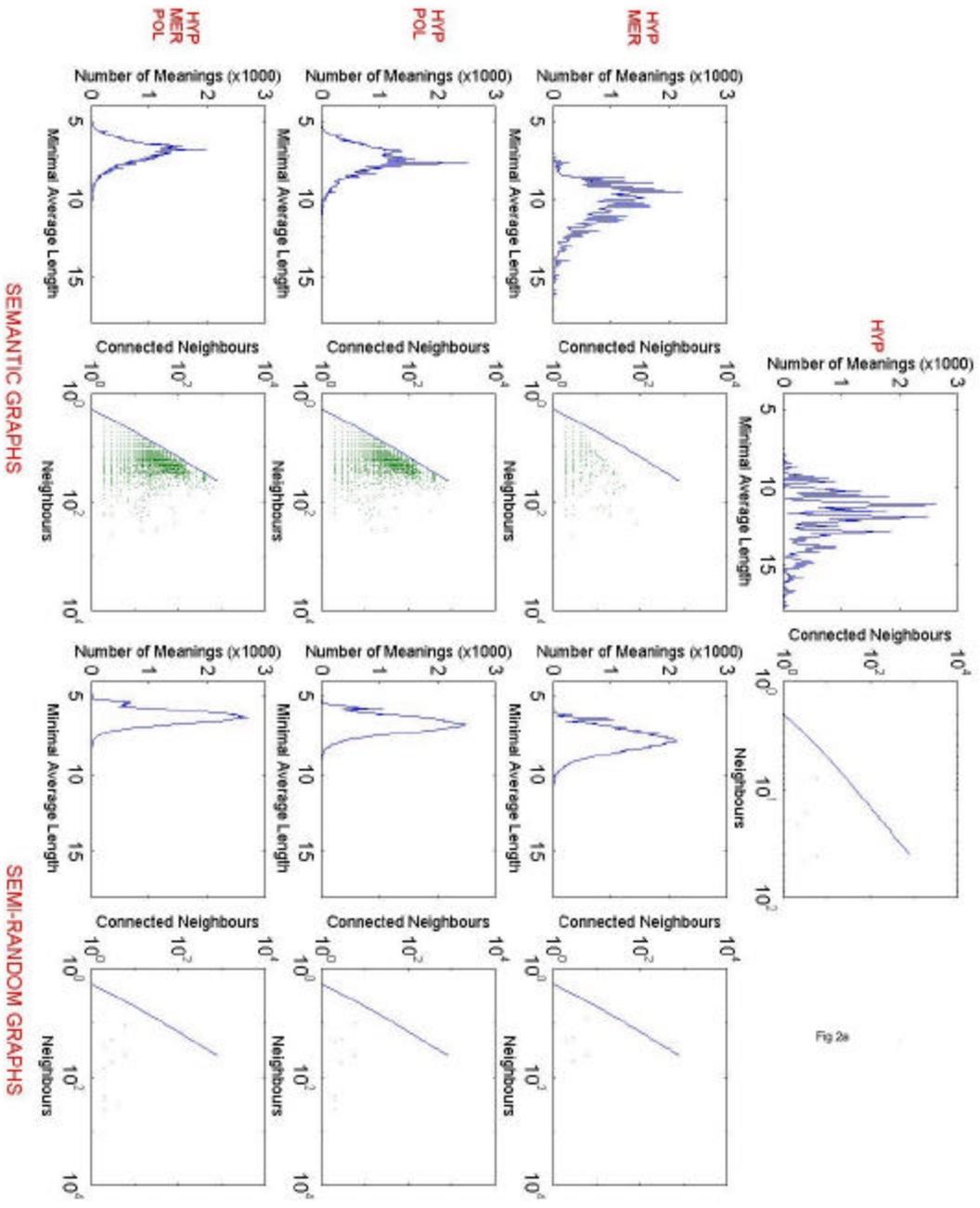

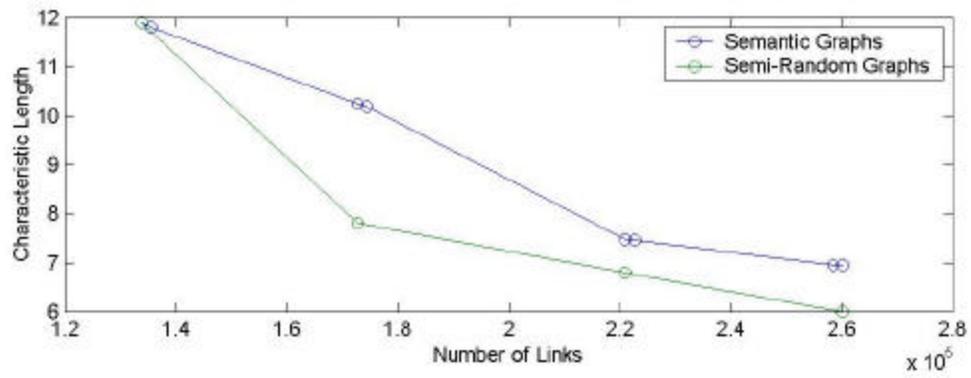

Fig 2b

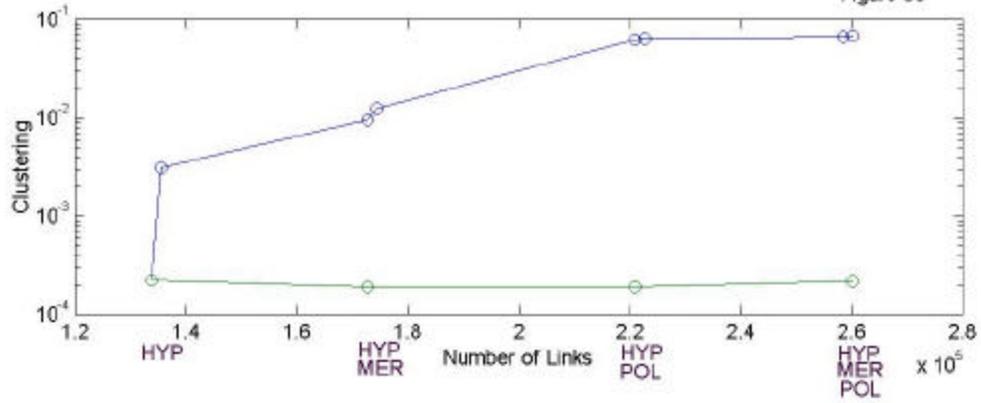

Figure 3c

HYP  HYP  HYP  HYP
     MER  POL  MER
               POL

Fig 3

○ Hypernymy-Polysemy (r=.545)
○ Hypernymy (r=.904)

Deepness vs. Average Minimal Length

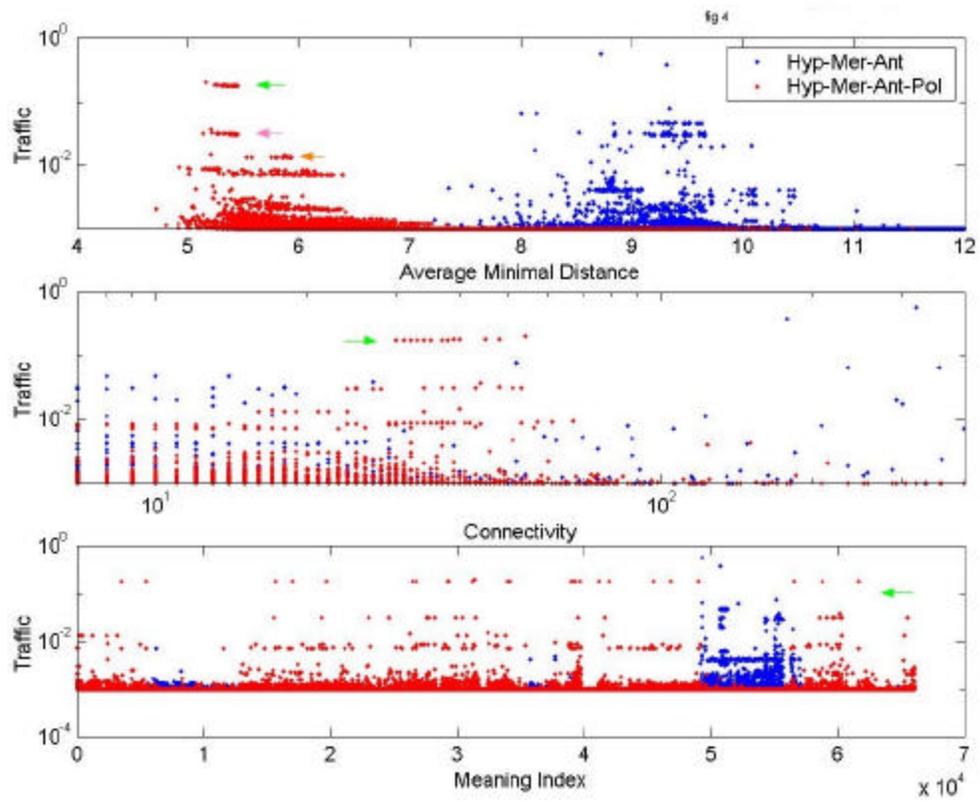